\documentclass[fleqn,10pt]{wlscirep}
\usepackage[utf8]{inputenc}
\usepackage[T1]{fontenc}

\title{Machine Learning Predicts Upper Secondary Education Dropout as Early as the End of Primary School}

\author[1,*]{Maria Psyridou}
\author[2]{Fabi Prezja}
\author[3]{Minna Torppa}
\author[3]{Marja-Kristiina Lerkkanen}
\author[3]{Anna-Maija Poikkeus}
\author[4]{Kati Vasalampi}
\affil[1]{Department of Psychology, University of Jyväskylä, 40014, Jyväskylä, Finland}
\affil[2]{Faculty of Information Technology, University of Jyväskylä, 40014, Jyväskylä, Finland}
\affil[3]{Department of Teacher Education, University of Jyväskylä, 40014, Jyväskylä, Finland}
\affil[4]{Department of Education, University of Jyväskylä, 40014, Jyväskylä, Finland}

\affil[*]{maria.m.psyridou@jyu.fi}

\begin{abstract}
Education plays a pivotal role in alleviating poverty, driving economic growth, and empowering individuals, thereby significantly influencing societal and personal development. However, the persistent issue of school dropout poses a significant challenge, with its effects extending beyond the individual. While previous research has employed machine learning for dropout classification, these studies often suffer from a short-term focus, relying on data collected only a few years into the study period. This study expanded the modeling horizon by utilizing a 13-year longitudinal dataset, encompassing data from kindergarten to Grade 9. Our methodology incorporated a comprehensive range of parameters, including students' academic and cognitive skills, motivation, behavior, well-being, and officially recorded dropout data. The machine learning models developed in this study demonstrated notable classification ability, achieving a mean area under the curve (AUC) of 0.61 with data up to Grade 6 and an improved AUC of 0.65 with data up to Grade 9. Further data collection and independent correlational and causal analyses are crucial. In future iterations, such models may have the potential to proactively support educators' processes and existing protocols for identifying at-risk students, thereby potentially aiding in the reinvention of student retention and success strategies and ultimately contributing to improved educational outcomes.

\end{abstract}
\begin{document}

\flushbottom
\maketitle
%
\thispagestyle{empty}

\section*{Introduction}

Education is often heralded as the key to poverty reduction, economic prosperity, and individual empowerment, and it plays a pivotal role in shaping societies and fostering individual growth\cite{huisman_keeping_2015, breton_can_2004, world_human_2021}. However, the specter of school dropout casts a long shadow, with repercussions extending far beyond the individual. Dropping out of school is not only a personal tragedy but also a societal concern; it leads to a lifetime of missed opportunities and reduced potential alongside broader social consequences, including increased poverty rates and reliance on public assistance. Existing literature has underscored the link between school dropout and diminished wages, unskilled labor market entry, criminal convictions, and early adulthood challenges, such as substance use and mental health problems\cite{backman_high_2017, bjerk_re-examining_2012, campolieti_labour_2010, dragone_high_2021}. The socioeconomic impacts, which range from reduced tax collections and heightened welfare costs to elevated healthcare and crime expenditures, signal the urgency of addressing this critical issue\cite{lee_societal_2011}. Therefore, understanding and preventing school dropout is crucial for both individual and societal advancement.

Beyond its economic impact, education differentiates individuals within the labor market and serves as a vehicle for social inclusion. Students’ abandonment of the pursuit of knowledge translates into social costs for society and profound personal losses. Dropping out during upper secondary education disrupts the transition to adulthood, impedes career integration, and compromises societal well-being\cite{freudenberg_reframing_2007}. The strong link between educational attainment and adult social status observed in Finland and globally\cite{kallio_cumulative_2016} underscores the importance of upper secondary education as a gateway to higher education and the labor market.

An increase in school drop-out rates in many European countries\cite{gubbels_risk_2019} is leading to growing pockets of marginalized young people. In the European Union (EU), 9.6\% of individuals between 18 and 24 years of age did not engage in education or training beyond the completion of lower secondary education\cite{eurostat_early_2021}. This disconcerting statistic raises alarms about the challenge of preventing early exits from the educational journey. Finnish statistics\cite{official_statistics_of_finland_osf_discontinuation_2022} highlight that 0.5\% of Finnish students drop out during lower secondary school, but this figure is considerably higher at the upper secondary level, with dropout rates of 13.3\% in vocational school and 3.6\% in general upper secondary school.  Amid this landscape, there is a clear and pressing need to not only support out-of-school youths and dropouts but also identify potential dropouts early on and prevent their potential disengagement. In view of the far-reaching consequences of school dropout for individuals and societies, social policy initiatives have rightly prioritized preventive interventions.

Machine learning has emerged as a transformative technology across numerous domains, particularly promising for its capabilities in utilizing large datasets and leveraging non-linear relationships. Within machine learning, deep learning\cite{lecun2015deep} has gained significant traction due to its ability to outperform traditional methods given larger data samples. Deep learning has played a significant role in advancements in fields such as medical computer vision \cite{esteva_dermatologist-level_2017, liu_comparison_2019,prezja2023synthesizing, prezja_deepfake_2022,prezja2023improving, topol_high-performance_2019 } and, more recently, in large foundation models\cite{wornow2023shaky, peng2023kosmos, livne2023nach0, luo2023biomedgpt}. Although machine learning methods have significantly transformed various disciplines, their application in education remains relatively unexplored \cite{bernardo_profiling_2023, bilal_role_2022}.

In education, only a handful of studies have harnessed machine learning to automatically classify between cases of students dropping out from upper secondary education or continuing in education. Previous research in this field has been constrained by short-term approaches. For instance, some studies have focused on collecting and analyzing data within the same academic year \cite{kruger_explainable_2023, sara_high-school_2015}. Others have restricted their data collection exclusively to the upper secondary education phase \cite{chung_dropout_2019, lee_machine_2019, sansone_beyond_2019}, while one study has expanded its dataset to include data collection of student traits across both lower and upper secondary school years\cite{aguiar_who_2015}. Only one previous study has focused on predicting dropout within the next three years following the collection of trait data \cite{colak_oz_school_2023}, and another study aimed at predictions within the next five years \cite{sorensen_big_2019}. However, the process of dropping out of school often begins in early school years and is marked by a gradual disengagement and
disassociation from education\cite{schoeneberger_longitudinal_2012, robert_balfanz_preventing_2007}. These findings suggest that current machine learning models might need to incorporate data that spans further back into the past. In this study we extended this time horizon by leveraging a 13-year longitudinal dataset, utilizing features from kindergarten up to Grade 9. In this study, we provide the first results for the automatic classification of upper secondary school dropout and non-dropout, using data available as early as the end of primary school.

\section*{Results}

\begin{figure}[ht!]
\centering
\includegraphics[scale= 0.35]{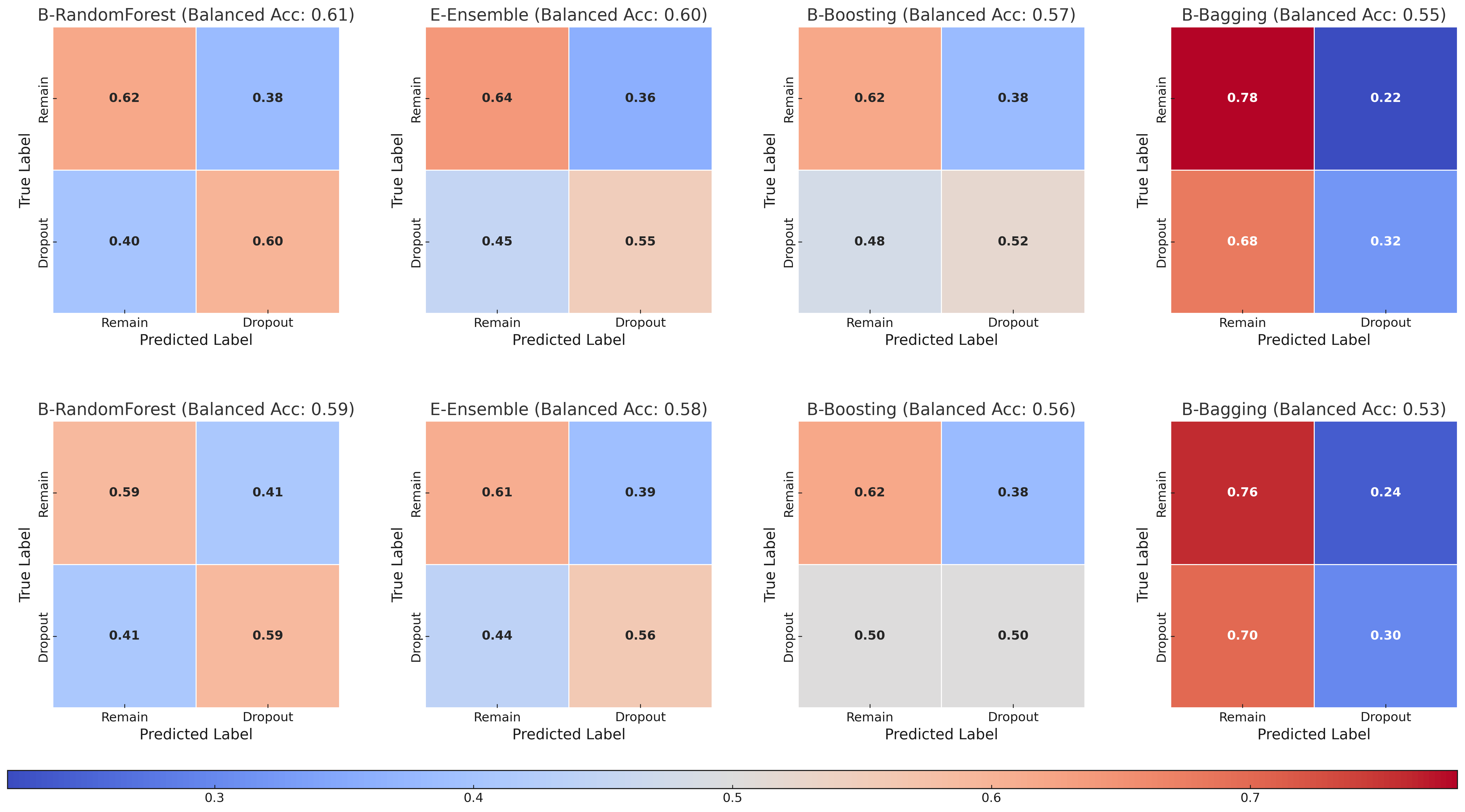}
\caption{Confusion matrices for classifiers using data up to Grade 9 (first row) and up to Grade 6 (second row) averaged across all folds in six-fold cross-validation.}
\label{fig:conf}
\end{figure}

This study utilized a comprehensive 13-year longitudinal dataset from kindergarten through upper secondary education. We applied machine learning techniques with data up to Grade 9, and subsequently with data up to Grade 6, to classify registered upper secondary education dropout and non-drop out status. The dataset included a broad range of educational data on students’ academic and cognitive skills, motivation, behavior, and well-being. Given the imbalance observed in the target, we trained four classifiers: Balanced Random Forest, or B-RandomForest; Easy Ensemble (AdaBoost Ensemble), or E-Ensemble; RSBoost (Adaboost), or B-Boosting; and Bagging Decision Tree, or B-Bagging. The performance of each classifier was evaluated using six-fold cross-validation, as shown in Fig. \ref{fig:conf} and Table \ref{table:classifiers_rounded}.

\begin{figure}[ht!]
\centering
\includegraphics[width=\linewidth]{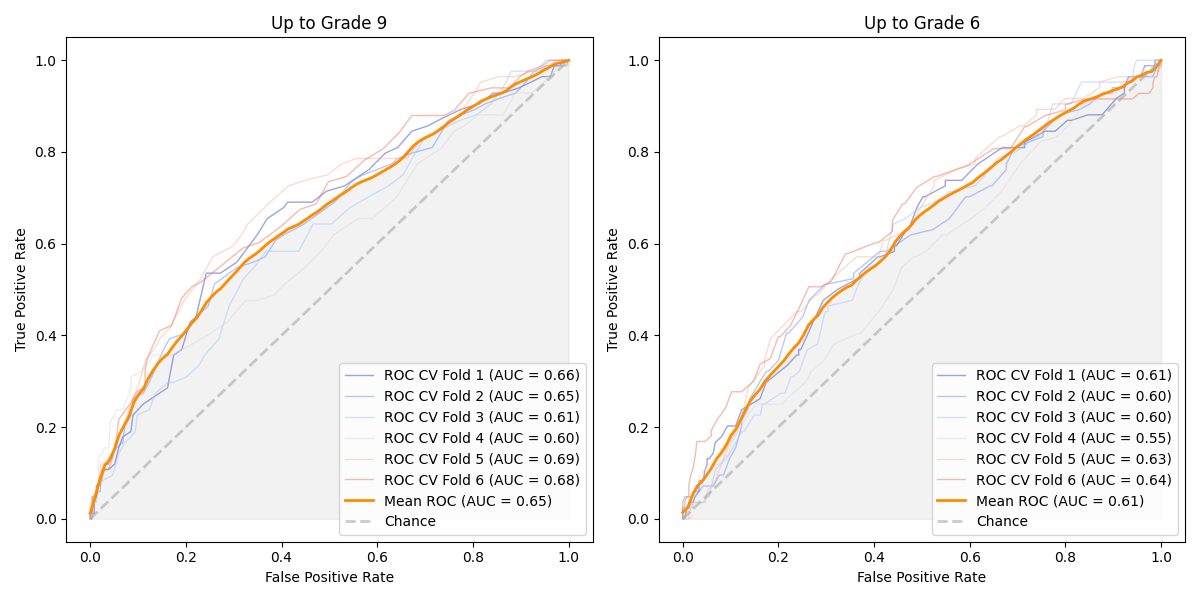}
\caption{The ROC Curves for the B-RandomForest classifiers from cross-validation. 
(a) Curve for the B-RandomForest classifier trained using data up to Grade 9. (b) Curve for another classifier instance trained using data up to Grade 6.}
\label{fig:rocs}
\end{figure}

Our analysis using data up to Grade 9 (Fig. \ref{fig:conf}, Table \ref{table:classifiers_rounded}), revealed that the B-RandomForest classifier was the most effective, as it achieved the highest balanced mean accuracy (0.61). It also showed a recall rate of 0.60 (i.e., dropout class) and a specificity of 0.62 (i.e., non-dropout class). While the other classifiers matched or exceeded the specificity (B-Bagging: 0.78, E-Ensemble: 0.64, B-Boosting: 0.62), they underperformed in classifying true positives (B-Bagging: 0.32, B-Boosting: 0.50, E-Ensemble: 0.56) and had higher false negative rates (B-Bagging: 0.68, B-Boosting: 0.48, E-Ensemble: 0.45). The B-RandomForest classifier demonstrated a mean area under the curve (AUC) of 0.65, which indicated good discriminative ability (Fig. \ref{fig:rocs}).

\begin{figure}[ht!]
\centering
\includegraphics[width=\linewidth]{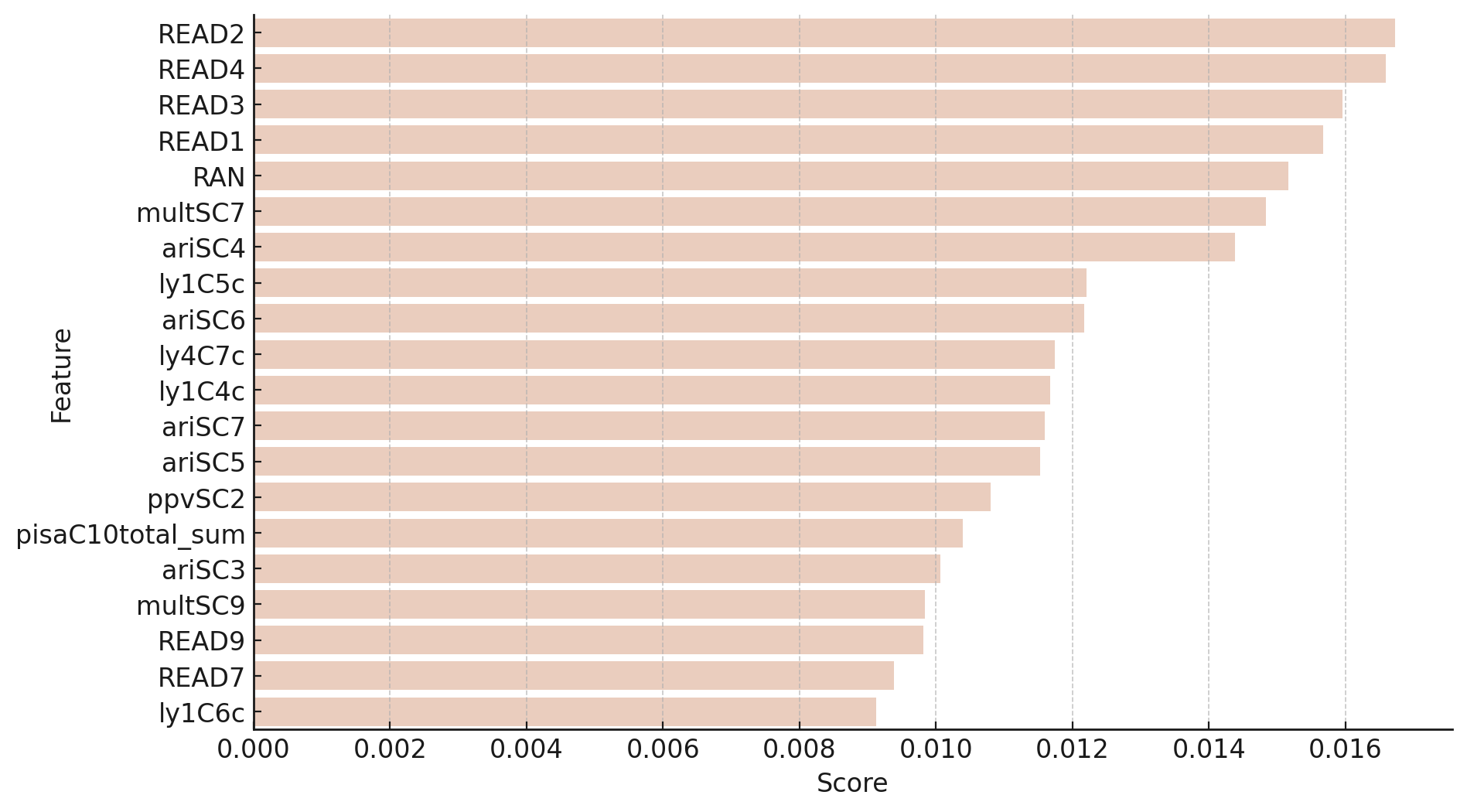}
\caption{The top ranked 20 features for the B-RandomForest using data up to Grade 9. Features are listed in order of average score from top to bottom. The scores are averages from across all folds of the six-fold cross-validation. The features listed pertain to:
READ2=Reading fluency, Grade 2; READ4=Reading fluency, Grade 4; READ3=Reading fluency, Grade 3; READ1=Reading fluency, Grade 1; RAN=Rapid Automatized Naming, Kindergarten; multSC7=Multiplication, Grade 4; ariSC4=Arithmetic, Grade 1 spring; ly1C5C=Reading comprehension, Grade 2; ariSC6=Arithmetic, Grade 3; ly4C7C=Reading comprehension, Grade 4; ly1C4C=Reading comprehension, Grade 1; ariSC7=Arithmetic, Grade 4; ariSC5=Arithmetic, Grade 2; ppvSC2=Vocabulary, Kindergarten; pisaC10total\_sum=PISA, Grade 9; ariSC3=Arithmetic, Grade 1 fall; multSC9=Multiplication, Grade 7; READ9=Reading fluency, Grade 9; READ7=Reading fluency, Grade 7; ly1C6C=Reading comprehension, Grade 3}
\label{fig:features9}
\end{figure}

We further obtained the feature scores for the B-RandomForest models across the six-fold cross-validation (Fig. \ref{fig:features9}; for the full list, refer to Supplementary Table S1). The top 20 rankings of the features (averaged across folds) fell into two domains: cognitive skills and academic outcomes. The Supplementary Table S3 provides a detailed description of all features. Academic outcomes appeared as the dominant domain and included reading fluency skills in Grades 1, 2, 3, 4, 7, and 9, reading comprehension in Grade 1, 2, 3, and 4, PISA reading comprehension outcomes, arithmetic skills in Grades 1, 2, 3, and 4, and multiplication skills in Grades 4 and 7. Among the top ranked features were two cognitive skills assessed in kindergarten: rapid automatized naming (RAN) which involved naming a series of visual stimuli consisting of pictures of objects (e.g., a ball, a house) as quickly as possible and vocabulary.

\begin{table}[ht!]
\centering
\begin{tabular}{|l|c|c|c|c|c|}
\hline
\textbf{Classifier}    & \textbf{Accuracy} & \textbf{Balanced Accuracy} & \textbf{Recall} & \textbf{Precision} & \textbf{F1-Score} \\ \hline
E-Ensemble             & 0.616             & 0.596                      & 0.555           & 0.580              & 0.572             \\ \hline
B-Boosting             & 0.596             & 0.571                      & 0.517           & 0.550              & 0.547             \\ \hline
B-Bagging              & 0.660             & 0.550                      & 0.318           & 0.541              & 0.539             \\ \hline
B-RandomForest         & 0.611             & 0.607                      & 0.599           & 0.582              & 0.573             \\ \hline
\end{tabular}
\caption{Average performance metrics across six-fold cross-validation (data up to Grade 9).}
\label{table:classifiers_rounded}
\end{table}

\subsection*{Classifying school dropout using data up to Grade 6}

Using data from kindergarten up to Grade 6, we retrained the same four classifiers on this condensed dataset and evaluated their performance using a six-fold cross-validation method (Fig. \ref{fig:conf}, Table \ref{table:classifiers_rounded}). The B-RandomForest classifier performed the highest, with 
a balanced mean accuracy of 0.59. It showed a recall rate of 0.59 (dropout class) and a specificity of 0.59 (non-dropout class). In comparison, the other classifiers had higher specificities (B-Bagging: 0.76, B-Boosting: 0.62, E-Ensemble: 0.61) but lower true positives (recall rates: B-Bagging: 0.30, B-Boosting: 0.50, E-Ensemble: 0.56) and exhibited higher false negative rates (B-Bagging: 0.70, B-Boosting: 0.50, E-Ensemble: 0.44). The B-RandomForest classifier demonstrated an AUC of 0.61 (Fig. \ref{fig:rocs}). The performance of this classifier was slightly lower but comparable to that of the classifier that used the more extensive dataset up to Grade 9.

\begin{table}[h!]
\centering
\begin{tabular}{|l|c|c|c|c|c|}
\hline
\textbf{Classifier}    & \textbf{Accuracy} & \textbf{Balanced Accuracy} & \textbf{Recall} & \textbf{Precision} & \textbf{F1-Score} \\ \hline
E-Ensemble             & 0.593             & 0.581                      & 0.557           & 0.564              & 0.552             \\ \hline
B-Boosting             & 0.587             & 0.561                      & 0.505           & 0.549              & 0.538             \\ \hline
B-Bagging              & 0.639             & 0.531                      & 0.302           & 0.532              & 0.531             \\ \hline
B-RandomForest         & 0.589             & 0.588                      & 0.587           & 0.569              & 0.554             \\ \hline
\end{tabular}
\caption{Average performance metrics across six-fold cross-validation (data up to Grade 6).}
\label{table:classifiers_updated}
\end{table}

\begin{figure}[ht!]
\centering
\includegraphics[width=\linewidth]{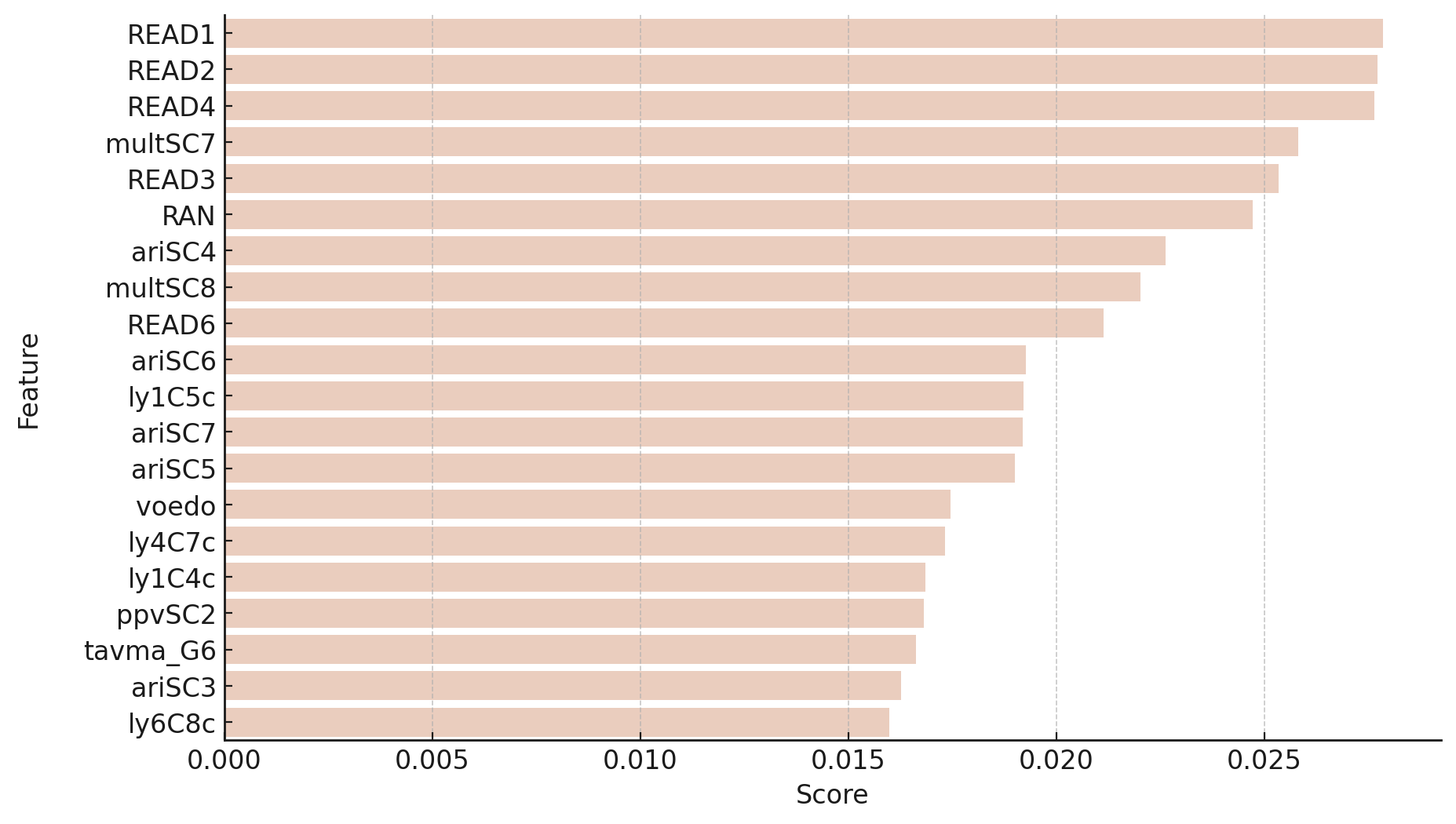}
\caption{The top ranked 20 features for the B-RandomForest using data up to Grade 6. Features are listed in order of average score from top to bottom. The scores are averages from across all folds of the six-fold cross-validation.
READ1=Reading fluency, Grade 1; READ2=Reading fluency, Grade 2; READ4=Reading fluency, Grade 4; multSC7=Multiplication, Grade 4; READ3=Reading fluency, Grade 3; RAN=Rapid Automatized Naming, Kindergarten; ariSC4=Arithmetic, Grade 1 spring; multSC8=Multiplication, Grade 6; READ6=Reading fluency, Grade 6; ariSC6=Arithmetic, Grade 3; ly1C5C=Reading comprehension, Grade 2; ariSC7=Arithmetic, Grade 4; ariSC5=Arithmetic, Grade 2; voedo=Parental education; ly4C7C=Reading comprehension, Grade 4; ly1C4C=Reading comprehension, Grade 1; ppvSC2=Vocabulary, Kindergarten; tavma\_g6=Task value for math, Grade 6; ariSC3=Arithmetic, Grade 1 fall; ly6C8C=Reading comprehension, Grade 6}
\label{fig:features6}
\end{figure}

We obtained the feature scores for the B-RandomForest models across the six-fold cross-validation with data up to Grade 6 (Fig. \ref{fig:features6}; for the full list, refer to Supplementary S2). The top 20 feature ranks included four domains: cognitive skills, academic outcomes, motivation, and family background. The Supplementary Information contains a detailed description of all features (Table S3). Similarly to the previous models academic outcomes ranked highest, consisting of reading fluency skills in Grades 1, 2, 3, 4, and 6, reading comprehension in Grades 1, 2, 4, and 6, arithmetic skills in Grades 1, 2, 3, and 4, and multiplication skills in Grades 4 and 6. Motivational factors, parental education level and two cognitive skills assessed in kindergarten – RAN and vocabulary – were also included in the ranking.

\section*{Discussion}

This study signifies a major advancement in educational research, as it provides the first predictive models leveraging data from as early as kindergarten to forecast upper secondary school dropout. By utilizing a comprehensive 13-year longitudinal dataset from kindergarten through upper secondary education, we developed predictive models using the Balanced Random Forest (B-RandomForest) classifier, which effectively predicted both dropout and non-dropout cases from as early as Grade 6.

The classifier's consistency was evident from its performance, which showed only a slight decrease in the AUC from 0.65 with data up to Grade 9 to 0.61 with data limited up to Grade 6. These results are particularly significant since they demonstrate predictive ability. Upon further validation and investigation, and by collecting more data, this approach may assist in the prediction of dropout and non-dropout as early as the end of primary school. However, it is important to note that the deployment and practical application of these findings must be preceded by further data collection, study, and validation. The developed predictive models offered some substantial indicators for future proactive approaches to help educators in their established protocols for identifying and supporting at-risk students. Such an approach could set a new precedent for enhancing student retention and success, potentially leading to transformative changes in educational systems and policies. While our predictive models marked a significant advancement in early automatic identification, it is important to recognize that this study is just the first step in a broader process.

The use of register data was a strength of this study because it allowed us to conceptualize dropout not merely as a singular event but as a comprehensive measure of on-time upper secondary education graduation. This approach is particularly relevant for students who do not graduate by the expected time, as it highlights their high risk of encountering problems in later education and the job market and underscores the need for targeted supplementary support\cite{knowles_needles_2015, rumberger_why_2012}. This conceptualization of dropout offers several advantages\cite{knowles_needles_2015} as it aligns with the nuanced nature of dropout and late graduation dynamics in educational practice. Additionally, it avoids mistakenly applying the dropout category to students who switch between secondary school tracks yet still graduate within the expected timeframe or drop out multiple times but ultimately graduate on time. From the perspective of the school system, delays in graduation incur substantial costs and necessitate intensive educational strategies. This nuanced understanding of dropout and non-dropout underpins the primary objective of our approach: to help empower educators with tools that can assist them in their evaluation of intervention needs.

In our study, we adopted a comprehensive approach to feature collection, acknowledging that the process of dropping out begins in early school years\cite{schoeneberger_longitudinal_2012} and evolves through protracted disengagement and disassociation from education\cite{robert_balfanz_preventing_2007}. With over 300 features covering a wide array of domains — such as family background, individual factors, behavior, motivation, engagement, bullying, health behavior, media usage, cognitive skills, and academic outcomes — our dataset presents a challenge typical of high-dimensional data: the curse of dimensionality.This phenomenon, where the volume of the feature space grows exponentially with the number of features, can lead to sparsity of data and make pattern recognition more complex.

To address these challenges, we employed machine learning classifiers like Random Forest, which are particularly adept at managing high-dimensional data. Random Forest inherently performs a form of feature selection, which is crucial in high-dimensional spaces, by building each tree from a random subset of features. This approach not only helps in addressing the risk of overfitting but also enhances the model's ability to identify intricate patterns in the data. This comprehensive analysis, with the use of machine learning, not only advances the methodology in automatic dropout and non-dropout prediction but also provides educators and policymakers with valuable tools and insights into the multifaceted nature of dropout and non-drop out identification from the perspective of machine learning classifiers.

In our study, the limited size of the positive class, namely the dropout cases, posed a significant challenge due to its impact on classification data balance. This imbalance steered our methodological decisions, leading us to forego both neural network synthesis and conventional oversampling techniques. Instead, we focused on using classification methods designed to handle highly imbalanced datasets. Our strategy was geared towards effectively addressing the issues inherent in working with severely imbalanced classification data.

Another important limitation to acknowledge pertains to the initial dataset and the subsequent handling of missing data. The study initially recruited around 2,000 kindergarten-age children and then invited their classmates to join the study at each subsequent educational stage. While this approach expanded the participant pool, it also resulted in a significant amount of missing data in many features. To maintain reliability, we excluded features with more than 30\% missing values. This aspect of our methodological approach highlights the challenges of managing large-scale longitudinal data. Future studies might explore alternative strategies for handling missing data or investigate ways to include a broader range of features for feature selection,  while mitigating the impact of incomplete data and the curse of dimensionality.

Despite these limitations, this study confronts the shortcomings of current research, particularly the focus on short-term horizons. Previous studies that have used machine learning to predict upper secondary education dropout have operated within limited timeframes – by collecting data on student traits and dropout cases within the same academic year\cite{kruger_explainable_2023, sara_high-school_2015}, limiting the collection of data on student traits to upper secondary education\cite{chung_dropout_2019, lee_machine_2019, sansone_beyond_2019}, and by collecting data on student traits during both lower and upper secondary school years\cite{aguiar_who_2015}. Two previous studies have focused on predicting dropout within three years\cite{colak_oz_school_2023} and five years \cite{sorensen_big_2019}, respectively, of collecting the data. The present study has extended this horizon by leveraging a 13-year longitudinal dataset, utilizing features from kindergarten, and predicting upper secondary school dropout and non-dropout as early as the end of primary school.

Our study identified a set of top features from Grades 1 to 4 that were highlighted by the Random Forest classifier as influential in predicting school dropout or non-dropout status. These features included aspects like reading fluency, reading comprehension, and arithmetic skills. These top feature rankings did not significantly change with data up to Grades 9 and 6. It is important to note that these features were identified based on their utility in improving the model's predictions within the dataset and cross-validation and should not be interpreted as causal or correlational factors for dropout and non-dropout rates. Given these limitations, and considering known across-time feature correlations \cite{aunola_developmental_2004, jessie_ricketts_reading_2020, verhoeven_prediction_2008,khanolainen_longitudinal_2020, psyridou_developmental_2023, psyridou_developmental_2021}, we find it pertinent to suggest further speculative discussions of this ranking consistency between early and later academic grades. If, upon further data collection, validation, and correlational and causal analysis this kind of ranking profile is re-established and validated, it could indicate that early proficiency in these key academic areas could potentially be an important factor influencing students’ educational trajectory and dropout risk.

In conclusion, this study represented a significant leap forward in educational research by developing predictive models that automatically distinguished between dropouts and non-dropouts as early as Grade 6. Utilizing a comprehensive 13-year longitudinal dataset, our research enriches existing knowledge of automatic school dropout and non-dropout detection and surpasses the time-frame confines of prior studies. While incorporating  data up to Grade 9 enhanced predictive accuracy, the primary aim of our study was to predict potential school dropout status at an early stage. The Balanced Random Forest classifier demonstrated proficiency across educational stages. Although confronted with challenges such as handling missing data and dealing with small positive class sizes, our methodological approach was meticulously designed to address such issues. 

The developed predictive models demonstrate potential for further investigation. Additional data and further validation using independent test sets are essential. Further independent correlational and causal analyses are also crucial. In future iterations, such models may have the potential to proactively support educators' processes and existing protocols for identifying at-risk students, thereby potentially aiding in the reinvention of student retention and success strategies, and ultimately contributing to improved educational outcomes.

\section*{Methods}

\begin{figure}[ht]
\centering
\includegraphics[scale=0.25]{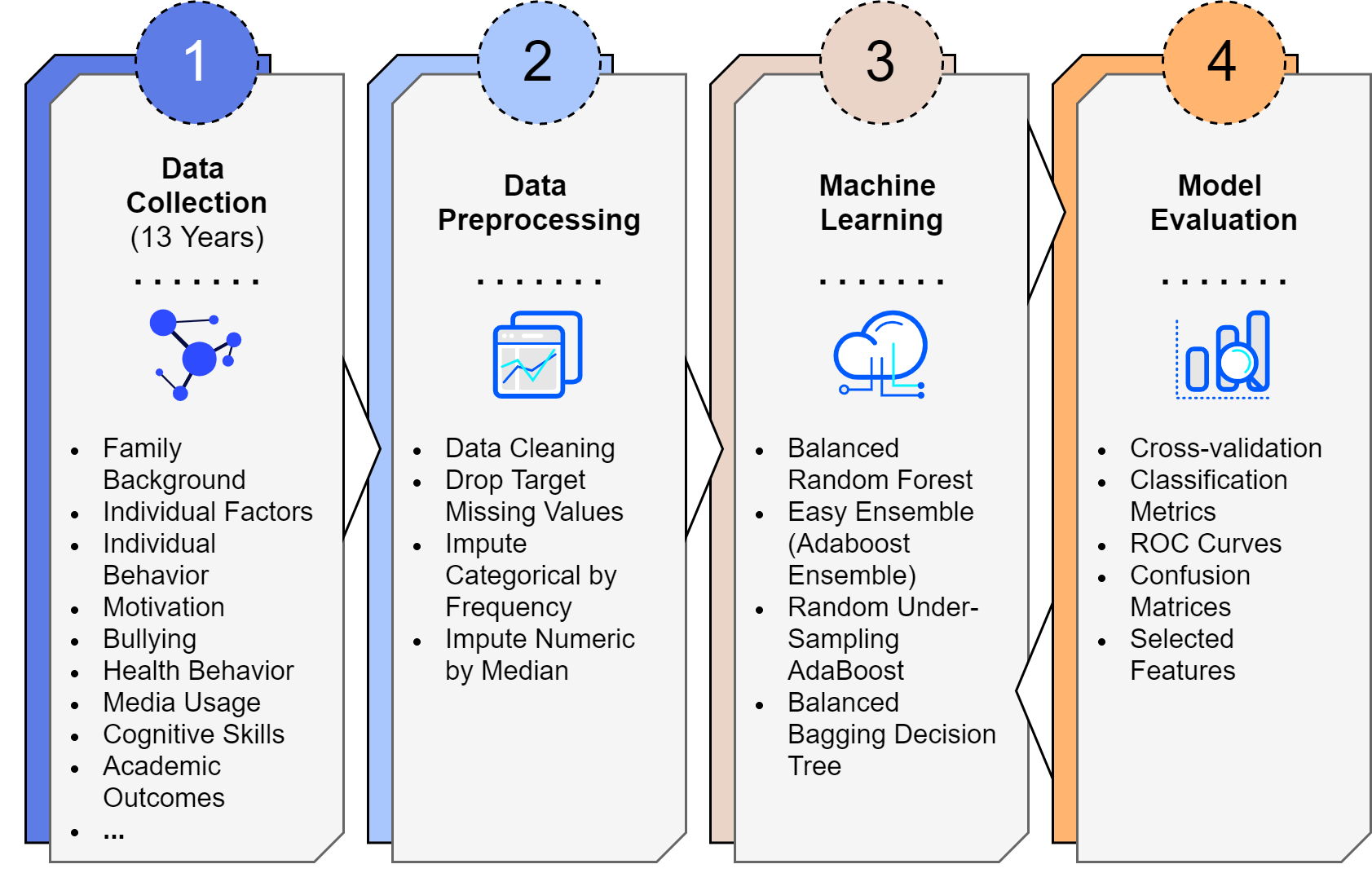}
\caption{Proposed research workflow. Our process begins with data collection over 13 years, from kindergarten to the end of upper secondary school (Step 1), followed by data processing which includes cleaning and imputing missing feature values (Step 2). We then apply four machine learning models for dropout and non-dropout classification (Step 3), and evaluate these models using 6-fold cross-validation, focusing on performance metrics and ROC curves (Step 4).}
\label{fig:flowchart}
\end{figure}

We trained and validated machine learning models, with a 13-year longitudinal dataset, to create classification models for upper secondary school dropout. Four supervised classification algorithms were utilized: Balanced Random Forest (B-RandomForest), Easy Ensemble (Adaboost Ensemble), RSBoost (Adaboost), and the Bagging Decision Tree. Six-fold cross-validation was used for the evaluation of performance. Confusion matrices were estimated for each model to evaluate their performance (Fig. \ref{fig:flowchart}).

\subsection*{Sampling}
This study used existing longitudinal data from the “First Steps” follow-up study\cite{Lerkkanen_2006} and its extension, the “School Path: From First Steps to Secondary and Higher Education” study\cite{Vasalampi_2016}. The entire follow-up spanned a 13-year period, from kindergarten to the third (final) year of upper secondary education. In the “First Steps” study, approximately 2,000 children born in 2000 were followed 10 times from kindergarten to the end of lower secondary school (Grade 9) in four municipalities around Finland (two medium-sized, one large, and one rural). The goal was to examine students’ learning, motivation, and problem behavior, including their academic performance, motivation and engagement, social skills, peer relations, and well-being, in different interpersonal contexts. The rate at which the contacted parents agreed to participate in the study ranged from 78\% to 89\% in the towns and municipalities – depending on the town or municipality. Ethnically and culturally, the sample was very homogeneous and representative of the Finnish population, and parental education levels were very close to the national distribution in Finland \cite{official_statistics_of_finland_2007}. In the “School Path” study, the participants of the “First Steps” follow-up study and their new classmates ($N = 4160$) were followed twice after the transition to upper secondary education at Grades 10 and 12.

The present study focused on those participants who took part in both the “First Steps” study and the “School Path” study. Data from three time points across three phases of the follow-up were used. Data collection for Time 1 (T1) took place in Fall 2006 and Spring 2007, when the participants entered kindergarten (age 6). Data collection for Time 2 (T2) took place during comprehensive school (ages 7–16), which extended from the beginning of primary school (Grade 1) in Fall 2007 to the end of the final year of the lower secondary school (Grade 9) in Spring 2016. For Time 3 (T3), data were collected at the end of 2019, 3.5 years after the start of upper secondary education. We focused on students who enrolled in either general upper secondary school (the academic track) or vocational school (the vocational track) following comprehensive school, as these tracks represent the choices available for young individuals in Finland. Common reasons for not completing school within 3.5 years included students deciding to discontinue their education or not fulfilling specific requirements (e.g., failing mandatory courses) during their schooling.

At T1 and T2, questionnaires were administered to the participants in their classrooms during normal school days, and their academic skills were assessed through group-administered tasks. Questionnaires were administered to parents as well. At T3, register information on the completion of upper secondary education was collected from school registers. In Finland, the typical duration of upper secondary education is three years. For the data collection in comprehensive school (T1 and T2), written informed consent was obtained from the participants’ guardians. In the secondary phase (T3), the participants themselves provided written informed consent to confirm their voluntary participation. The ethical statements for the follow-up study were obtained in 2006 and 2018 from the Ethical Committee of the University of Jyväskylä.

\subsection*{Measures}
\begin{figure}[ht!]
\centering
\includegraphics[width=\linewidth]{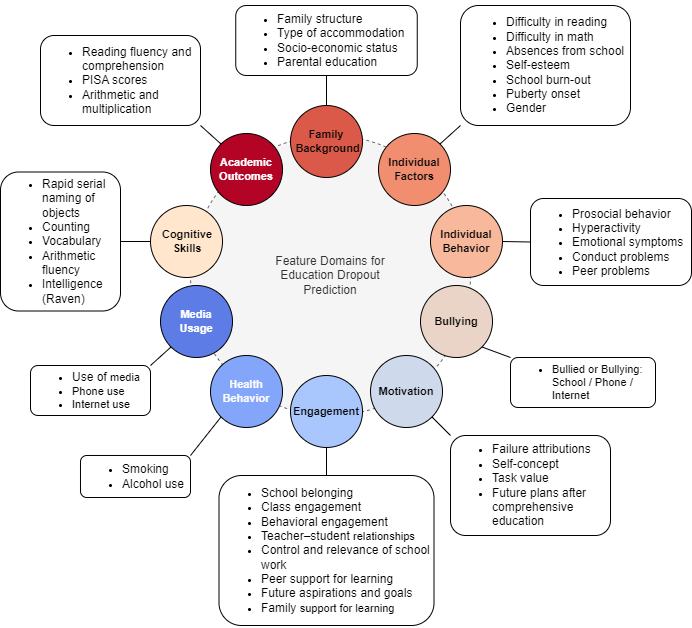}
\caption{Features domains used for the classification of education dropout and non-dropout. The model incorporated a set of 311 features, categorized into 10 domains: family background, individual factors, behavior, motivation, engagement, bullying experiences, health behavior, media usage, cognitive skills, and academic outcomes. Each domain encompassed a variety of measures.}
\label{fig:variables}
\end{figure}

The target variable in the 13-year follow-up was the participant’s status 3.5 years after starting upper secondary education, as determined from the school registers. Participants who had not completed upper secondary education by this time were coded as having dropped out. Initially, we considered the assessment of 586 features. However, as is common in longitudinal studies, missing values were identified in all of them. Features with more than 30\% missing data were excluded from the analysis, and a total of 311 features were used (with one-hot encoding) (see Supplementary Table S3). These features covered family background (e.g., parental education, socio-economic status), individual factors (e.g., gender, absences from school, school burn-out), the individual’s behavior (e.g., prosocial behavior, hyperactivity), motivation (e.g., self-concept, task value), engagement (e.g., teacher-student relationships, class engagement), bullying (e.g., bullied, bullying), health behavior (e.g., smoking, alcohol use), media usage (e.g., use of media, phone, internet), cognitive skills (e.g., rapid naming, raven), and academic outcomes (i.e., reading fluency, reading comprehension, PISA scores, arithmetic, and multiplication).  Figure \ref{fig:variables} presents an overview of the features used. The Supplementary Table S3 provides details about the features included.

\subsection*{Data Processing}
In our study, we employed a systematic approach to address missing values in the dataset. Initially, the percentage of missing data was calculated for each feature, and features exhibiting more than 30\% missing values were excluded. For categorical features, imputation was performed using the most frequent value within each feature, while a median-based strategy was applied to numeric features. To ensure unbiased imputation, imputation values were derived from a temporary dataset where the majority class (i.e., non-dropout cases) was randomly sampled to match the size of the positive class (i.e., dropout cases).

\subsection*{Machine Learning}

In our study, we utilized a range of balanced classifiers from the Imbalanced Learning Python package\cite{JMLR:v18:16-365} for benchmarking. These classifiers were employed with their default hyperparameter settings. Our selection included Balanced Random Forest, Easy Ensemble (Adaboost Ensemble), RSBoost (Adaboost), and Bagging Decision Tree. Notably, the Balanced Random Forest classifier played a pivotal role in our study. We delve into its performance, specific configuration, and effectiveness in the following section. Below are descriptions of each classifier:

\begin{enumerate}
    \item \textbf{Balanced Random Forest}: This classifier modifies the traditional random forest\cite{breiman2001random} approach by randomly under-sampling each bootstrap sample to achieve balance. In our study, we refer to the classifier as "B-RandomForest".

    \item \textbf{Easy Ensemble (Adaboost Ensemble)}: This classifier, known as EasyEnsemble\cite{liu2008exploratory}, is a collection of AdaBoost\cite{freund1997decision} learners that are trained on differently balanced bootstrap samples. The balancing is realized through random under-sampling. In our study, we refer to the classifier as "E-Ensemble".
    
    \item \textbf{RSBoost (Adaboost) }: This classifier integrates random under-sampling into the learning process of AdaBoost. It under-samples the sample at each iteration of the boosting algorithm. In our study, we refer to the classifier as "B-Boosting".

    \item \textbf{Bagging Decision Tree}: This classifier operates similarly to the standard Bagging\cite{breiman1996bagging} classifier in the scikit-learn library\cite{scikit-learn} using decision trees\cite{quinlan1986induction}, but it incorporates an additional step to balance the training set by using a sampler. In our study, we refer to the classifier as "B-Bagging".
    
\end{enumerate}

Each of these classifiers was selected for their specific strengths in handling class imbalances, a critical consideration of our study's methodology. The next section elaborates on the performance and configurations of these classifiers, particularly B-RandomForest.

\subsubsection*{Random Forest}
\newcommand{\qvec}[1]{\textbf{\textit{#1}}}

The Random Forest (RF) method, introduced by Breiman in 2001 \cite{breiman2001random}, is a machine learning approach that employs a collection of decision trees for prediction tasks. This method's strength lies in its ensemble nature, where multiple "weak learners" (individual decision trees) combine to form a "strong learner" (the RF). Typically, decision trees in an RF make binary predictions based on various feature thresholds. The mathematical representation of a single decision tree's prediction,  ($T_d$) for an input vector $\qvec{I}$ is given by the following formula:

\begin{equation}\label{eq:tree}
T_d(\qvec{I}) = \sum_{i=1}^{n} v_i\delta(f_i(\qvec{I}) < t_i)
\end{equation}

Here, $n$ signifies the total nodes in the tree, $v_i$ is the value predicted at the $i$-th node, $f_i(\qvec{I})$ is the $i$-th feature of the input vector $\qvec{I}$, $t_i$ stands for the threshold at the $i$-th node, and $\delta$ represents the indicator function.

In an RF, the collective predictions from $D$ individual decision trees are aggregated to form the final output. For regression problems, these outputs are typically averaged, whereas a majority vote (mode) approach is used for classification tasks. The prediction formula for an RF ($F_D$) on an input vector $\qvec{I}$, is as follows:

\begin{equation}\label{eq:forest}
F_D(\qvec{I}) = \frac{1}{D} \sum_{d=1}^{D} T_d(\qvec{I})
\end{equation}

In this equation, $T_d(\qvec{I})$ is the result from the $d$-th tree for input vector $\qvec{I}$, and $D$ is the count of decision trees within the forest. Random Forests are particularly effective for reducing overfitting compared to individual decision trees because they average results across a plethora of trees. In our study, we utilized 100 estimators with default settings from the scikit-learn library \cite{scikit-learn}.

\subsection*{Figures of Merit}

To evaluate the efficacy of our classification models, we employed a set of essential evaluative metrics, known as figures of merit.

The accuracy metric reflects the fraction of correct predictions (encompassing both true positive and true negative outcomes) in comparison to the overall number of predictions. The formula for accuracy is as follows:

\begin{equation}
\mathrm{Accuracy} = \frac{\mathrm{TP} + \mathrm{TN}}{\mathrm{TP} + \mathrm{TN} + \mathrm{FP} + \mathrm{FN}}
\end{equation}

Notably, given the balanced nature of our target data, the accuracy rate in our analysis equated to the definition of balanced accuracy.

Precision, or the positive predictive value, represents the proportion of true positive predictions out of all positive predictions made. The equation to determine precision is as follows:

\begin{equation}
\mathrm{Precision} = \frac{\mathrm{TP}}{\mathrm{TP} + \mathrm{FP}}
\end{equation}

Recall, which is alternatively called sensitivity, quantifies the percentage of actual positives that were correctly identified. The formula for calculating recall is as follows:

\begin{equation}
\mathrm{Recall} = \frac{\mathrm{TP}}{\mathrm{TP} + \mathrm{FN}}
\end{equation}

Specificity, also known as the true negative rate, measures the proportion of actual negatives that were correctly identified. The formula for specificity is as follows:

\begin{equation}
\mathrm{Specificity} = \frac{\mathrm{TN}}{\mathrm{TN} + \mathrm{FP}}
\end{equation}

The F1 Score is the harmonic mean of precision and recall, providing a balance between the two metrics. It is particularly useful when the class distribution is imbalanced. The formula for the F1 Score is as follows:

\begin{equation}
\mathrm{F1\ Score} = 2 \cdot \frac{\mathrm{Precision} \cdot \mathrm{Recall}}{\mathrm{Precision} + \mathrm{Recall}}
\end{equation}

In these formulas, \(\mathrm{TP}\) represents true positives, \(\mathrm{TN}\) stands for true negatives, \(\mathrm{FP}\) refers to false positives, and \(\mathrm{FN}\) denotes false negatives. 

The balanced accuracy metric, as referenced by Brodersen et al. in 2010\cite{brodersen2010balanced}, is a crucial measure in the context of classification tasks, particularly when dealing with imbalanced datasets. This metric is calculated as follows:

\begin{equation}\label{eq:balacc}
BalAcc = \frac{1}{2}\left(\frac{TP}{TP+FN}+\frac{TN}{TN+FP}\right)
\end{equation}

Essentially, this equation is an average of the recall computed for each class. The balanced accuracy metric is particularly effective since it accounts for class imbalance by applying balanced sample weights. In situations where the class weights are equal, this metric is directly analogous to the conventional accuracy metric. However, when class weights differ, the metric adjusts accordingly and weights each sample based on the true class prevalence ratio. This adjustment makes the balanced accuracy metric a more robust and reliable measure in scenarios where the class distribution is uneven. In line with this approach, we also employed the macro average of F1 and Precision in our computations.

A confusion matrix is a vital tool for understanding the performance of a classification model. In the context of our study, the performance of each classification model was encapsulated by binary confusion matrices. One matrix was a \(2\times2\) table categorizing the predictions into four distinct outcomes. In the columns of the matrix,the classifications predicted by the model are represented and categorized as Predicted Positive and Predicted Negative. The rows signify the actual classifications, which are labeled as Actual Positive and Actual Negative.

\begin{itemize}
  \item The upper-left cell is the True Negatives (TN), which are instances where the model correctly predicted the negative class.
  \item The upper-right cell is the False Positives (FP), which are cases where the model incorrectly predicted the positive class for actual negatives.
  \item The lower-left cell is the False Negatives (FN), where the model incorrectly predicted the negative class for actual positives.
  \item Finally, the lower-right cell shows 'True Positives (TP)', where the model correctly predicted the positive class.
\end{itemize}

In our study, we aggregated the results from all iterations of the cross-validation process to generate normalized average binary confusion matrices. Normalization of the confusion matrix involves converting the raw counts of true positives, false positives, true negatives, and false negatives into proportions, which account for the varying class distributions. This approach allows for a more comparable and intuitive understanding of the model's performance, especially when dealing with imbalanced datasets. By analyzing the normalized matrices, we obtain a comprehensive view of the model's predictive performance across the entire cross-validation run, instead of relying on a single instance.

\subsection*{AUC Score}

The AUC score is a widely used metric in machine learning for evaluating the performance of binary classification models. Derived from the receiver operating characteristic (ROC) curve, the AUC score quantifies a model's ability to distinguish between two classes. The ROC curve plots the true positive rate (TPR) against the false positive rate (FPR) at various threshold settings. By varying the threshold that determines the classification decision, the ROC curve illustrates the trade-off between sensitivity (TPR) and specificity (1 – FPR). The TPR and FPR are defined as follows:

\begin{equation}
\mathrm{TPR} = \frac{\mathrm{TP}}{\mathrm{TP} + \mathrm{FN}}
\end{equation}

\begin{equation}
\mathrm{FPR} = \frac{\mathrm{FP}}{\mathrm{FP} + \mathrm{TN}}
\end{equation}

The AUC score represents the area under the ROC curve and ranges from 0 to 1. An AUC score of 0.50 is equivalent to random guessing and indicates that the model has no discriminative ability. On the other hand, a model with an AUC score of 1.0 demonstrates perfect classification. A higher AUC score suggests a better model performance in terms of distinguishing between the positive and negative classes.

\subsection*{Cross-validation}

In this study, we employed the stratified K-fold cross-validation method with $K=6$ to ascertain the robustness and generalizability of our approach \cite{kohavi1995study}. This method partitions the dataset into $k$ distinct subsets, or folds with an even distribution of class labels in each fold to reflect the overall dataset composition. For each iteration of the process, one of these folds is designated as the test set, while the remaining folds collectively form the training set. This cycle is iterated $k$ times, with a different fold used as the test set each time. This technique was crucial in our study to ensure that the model's performance would be consistently evaluated against varied data samples. One formal representation of this process with $K=6$, is as follows:

\begin{equation}
\mathrm{CV}(\mathcal{M}, \mathcal{D}) = \frac{1}{K} \sum_{k=1}^{K} \mathrm{Eval}(\mathcal{M}, \mathcal{D}_k^\mathrm{train}, \mathcal{D}_k^\mathrm{test})
\end{equation}

Here, \(\mathcal{M}\) represents the machine learning model, \(\mathcal{D}\) is the dataset, \(\mathcal{D}_k^\mathrm{train}\) and \(\mathcal{D}_k^\mathrm{test}\) respectively denote the training and test datasets for the \( k \)-th fold, and \(\mathrm{Eval}\) is the evaluation function (e.g., accuracy, precision, recall).Our AUC plots have been generated using the forthcoming version of utility functions from the Deep Fast Vision Python Library \cite{prezja2023deep}.

\section*{Data availability}
The datasets generated during and/or analysed during the current study are available from the corresponding author on reasonable request.

\bibliography{main}

\section*{Acknowledgements}
The First Steps Study was funded by by grants from the Academy of Finland (Grant numbers: 213486, 263891, 268586, 292466, 276239, 284439, and 313768). The School Path study was funded by grants from Academy of Finland (Grant numbers: 299506 and 323773).This research was also partly funded by the Strategic Research Council (SRC) established within the Academy of Finland (Grant numbers: 335625, 335727, 345196, 358490, and 358250 for the project CRITICAL and Grant numbers: 352648, 353392 for the project Right to Belong). In addition, Maria Psyridou was supported by the Academy of Finland (Grant number: 339418)

\section*{Author contributions statement}

M.P. conceived the experiment, was involved in data curation, and analysed the results. F.P. was involved in data curation and analysed the results. 
M.K.L. was involved in data collection. A.M.P. was involved in data collection. M.T. conceived the experiment, was involved in data curation and data collection. K.V. conceived the experiment and was involved in data curation and data collection. All authors reviewed the manuscript.

\section*{Additional information}
 \textbf{Competing interests}
 All authors declare that they have no conflicts of interest.

\end{document}